\documentstyle[aps,preprint,eqsecnum,floats,epsf]{revtex}
\tighten \draft
\begin{document}

\preprint{} \title{Completeness of Coherent
  States Associated with Self-Similar Potentials and Ramanujan's
  Integral Extension of the Beta Function}

\author{A.~N.~F. Aleixo\thanks{Electronic address: {\tt
      armando@if.ufrj.br}}} 
\address{Instituto de
F\'{\i}sica,  Universidade Federal  do Rio de Janeiro, RJ -  Brazil}

\author{A.~B. Balantekin\thanks{Electronic address: {\tt
      baha@nucth.physics.wisc.edu}}} 
\address{Department of Physics, University of Wisconsin;\\  Madison,
Wisconsin 53706 USA}

\author{M.~A. C\^andido Ribeiro\thanks{Electronic address: {\tt
macr@df.ibilce.unesp.br}},}  \address{Departamento de F\'{\i}sica -
Instituto de Bioci\^encias, Letras e Ci\^encias Exatas\\ UNESP, S\~ao
Jos\'e do Rio Preto, SP - Brazil}

\maketitle

\begin{abstract}
 A decomposition of identity is given as a complex integral over 
 the coherent states associated with a class of shape-invariant
 self-similar potentials. There is a remarkable connection between
 these coherent states and Ramanujan's integral extension of the beta
 function. 
\end{abstract}

\newpage
\section{Introduction}

Supersymmetric quantum mechanics is the study of pairs of Hamiltonians
with identical energy spectra and with eigenstates that are different,
but can be transformed into each other
\cite{Witten:nf,Cooper:1994eh}. Some, but not all, such pairs of
Hamiltonians share an integrability condition called shape-invariance
\cite{Gendenshtein:vs}. In Ref. \cite{Balantekin:1997mg} it was shown
that the shape-invariance condition has an underlying algebraic
structure and the associated Lie algebras were identified. Utilizing
this algebraic structure a general definition of coherent states for
shape-invariant potentials were introduced by several authors
\cite{Fukui:1993xv,Balantekin:1998wj}. These coherent states are
eigenstates of the annihilation operator of the shape-invariant
system. When the shape-invariant system is taken to be the standard
harmonic oscillator, q-analogue harmonic oscillator or
$SU(1,1)$-covariant system those coherent states reduce to the
well-known harmonic oscillator coherent states, q-analogue coherent
states or the Barut-Girardello $SU(1,1)$ coherent states
\cite{Barut:1970qf}, respectively.

A decomposition of the identity operator using q-integration is
available in the literature \cite{Gray:ea}. In this paper we use
shape-invariant algebraic techniques to obtain a decomposition of the
identity as a regular integral over complex variables. It turns out
that such an integral is very closely related to an integral
evaluated by Ramanujan in studying integral expansion of the beta
function. 

In the next section we review the relevant formulae for supersymmetric
quantum mechanics and the shape-invariance. In Section 3 we introduce
generalized coherent states for shape-invariant systems. In section 4
we show that, when the appropriate shape-invariant system is chosen, 
these reduce to the q-analogue coherent states. The proof
of completeness of these coherent states in terms of the Ramanujan's
integral is given in Section 5. Several brief remarks in Section 6
conclude the paper.  

\section{Supersymmetric Quantum Mechanics and Shape-Invariance}

In one-dimension supersymmetric quantum mechanics uses the operators
\cite{Witten:nf} 
\begin{equation}
\hat A \equiv W(x) + \frac{i}{\sqrt{2m}}\hat p\,,
\label{opa}
\end{equation}
\begin{equation}
\hat A^\dagger \equiv W(x) - \frac{i}{\sqrt{2m}}\hat p\,,
\label{opad}
\end{equation}
to write down the Hamiltonian in the form
\begin{equation}
\hat H - E_0 = \hat A^\dagger \hat A\,.
\label{eqhe}
\end{equation}
Here $E_0$ is the energy of the ground-state the wave-function of
which is annihilated by the operator $\hat A$:
\begin{equation}
\hat A\mid \Psi_0\rangle = 0 .
\label{eqgs}
\end{equation}
To illustrate the underlying algebraic structure the shape-invariance 
condition can be written in terms of the operators defined in
Eqs. (\ref{opa}) and (\ref{opad}) as \cite{Balantekin:1997mg} 
\begin{equation}
\hat A(a_1) \hat A^\dagger(a_1) =\hat A^\dagger (a_2) 
\hat A(a_2) + R(a_1) \,,
\label{eqsi}
\end{equation}
where $a_{1,2}$ are a set of parameters. The parameter $a_2$ is a
function of $a_1$ and the remainder $R(a_1)$ is independent of $\hat
x$ and $\hat p$. Introducing the similarity
transformation that replaces $a_1$ with $a_2$ in a given operator
\begin{equation}
\hat T(a_1)\, \hat O(a_1)\, \hat T^\dagger(a_1) = \hat O(a_2)
\label{eqsio}
\end{equation}
and the operators
\begin{equation}
\hat B_+ =  \hat A^\dagger(a_1)\hat T(a_1)
\label{eqba}
\end{equation}
\begin{equation}
\hat B_- =\hat B_+^\dagger =  \hat T^\dagger(a_1)\hat A(a_1)\,,
\label{eqbe}
\end{equation}
the Hamiltonian takes the form
\begin{equation}
\hat H - E_0 =\hat B_+\hat B_-\, .
\label{eqhb}
\end{equation}
Using Eqs. (\ref{eqsio}) through (\ref{eqbe}) one can show that  the
commutation relations
\begin{equation}
[\hat B_-,\hat B_+] =  \hat T^\dagger(a_1)R(a_1)\hat T(a_1) 
\equiv R(a_0)\,,
\label{eqcb1}
\end{equation}
and
\begin{equation}
[\hat B_+,R(a_0)] =  [R(a_1)-R(a_0)]\hat B_+\,,
\label{eqcb2}
\end{equation}
\begin{equation}
[\hat B_+,\{R(a_1)-R(a_0)\}\hat B_+] =
\{[R(a_2)-R(a_1)]-[R(a_1)-R(a_0)]\}\hat B_+^2 \,,
\label{eqcb3}
\end{equation}
and the Hermitian conjugates of the relations given in
Eq.~(\ref{eqcb2}) and Eq.~(\ref{eqcb3}) are satisfied
\cite{Balantekin:1997mg}. In the most general case the resulting Lie
algebra is infinite-dimensional. 

One class of shape-invariant potentials are 
reflectionless potentials with an infinite number of bound states,
also called self-similar potentials \cite{shabat,Spiridonov:md}.
Shape-invariance of such potentials were studied in detail in Refs. 
\cite{Khare:gg,Barclay:1993kt}. For such potentials the parameters are
related by a scaling:
\begin{equation}
a_n = q^{n-1}a_1\,.
\label{eqsc}
\end{equation}
For the simplest case studied in Ref. \cite{Barclay:1993kt} the
remainder of Eq.~(\ref{eqsi}) is given by
\begin{equation}
R(a_1)= ca_1 \,,
\label{eqsc1}
\end{equation}
where $c$ is a constant and the operator introduced in
Eq.~(\ref{eqsio}) by 
\begin{equation}
\hat T(a_1) = 
\exp{\left\{(\log q)a_1\frac{\partial}{\partial a_1}\right\}}\,.
\label{eqta}
\end{equation}
For this shape-invariant potential one can show that the scaled
operators 
\begin{equation}
\hat S_+ = \sqrt{q}\hat B_+R(a_1)^{-1/2}
\label{eqsma}
\end{equation}
and
\begin{equation}
\hat S_- = (\hat S_+)^\dagger = \sqrt{q} R(a_1)^{-1/2}\hat B_-\,,
\label{eqsme}
\end{equation}
satisfy the standard $q$-deformed oscillator relation
\cite{Arik:1973vg}:  
\begin{equation}
\hat S_-\hat S_+ - q\hat S_+\hat S_- = 1\,.
\label{eqcsq}
\end{equation}
The Hamiltonian takes the form
\begin{equation}
  \label{a1}
  \hat H - E_0 = R(a_1) \hat S_+ \hat S_- \, .
\end{equation}
The energy eigenvalues are
\begin{equation}
  \label{a11}
  E_n = R(a_1) \frac{1-q^n}{1-q} \,,
\end{equation}
and the normalized eigenstates are given by
\begin{equation}
  \label{a2}
  \mid  n\rangle = \sqrt{\frac{(1-q)^n}{(q;q)_n}} (\hat S_+)^n \mid
  0\rangle 
\end{equation}
where the $q$-shifted factorial $(q;q)_n$ is defined as $(z;q)_0 = 1$
and $(z;q)_n = \prod_{j=0}^{n-1}(1-zq^j)\,,$ $n = 1,2,\dots$.  

One should point out q-generalization of not only the standard
harmonic oscillator, but also of other exactly solvable problems are
available in the literature (see
e.g. \cite{Dayi:1994dr}). Shape-invariance properties of such
generalizations are yet unexplored. 

\section{Coherent States}

Coherent states for shape-invariant potentials were introduced in
Refs. \cite{Fukui:1993xv} and \cite{Balantekin:1998wj}. Here we follow
the notation of Ref. \cite{Balantekin:1998wj}. Using the right inverse
of $\hat B_-$
\begin{equation}
\hat H^{-1}\hat B_+ =  \hat B_-^{-1}, \>\>\>\> (\hat B_-\hat B_-^{-1}
=  1)  
\label{eqhb1}
\end{equation}
the coherent state for shape-invariant potentials with an infinite
number of energy eigenstates was defined in
Ref. \cite{Balantekin:1998wj} as
\begin{eqnarray}
\mid z\rangle_c &=& \mid 0\rangle\ + z\hat B_-^{-1}\mid 0\rangle +
z^2\hat B_-^{-2}\mid 0\rangle + \dots \nonumber \\
&=& \frac{1}{1-z\hat B_-^{-1}}\mid 0\rangle ,
\label{eqcs}
\end{eqnarray}
where $\mid 0\rangle$ is the ground state of the Hamiltonian in
Eq. (\ref{eqhb}). The coherent state in Eq. (\ref{eqcs}) is an
eigenstate of the operator $\hat B_-$: 
\begin{equation}
\hat B_-\mid z\rangle_c = z\mid z\rangle_c \>.  
\label{eqbcs}
\end{equation}

In this paper we use a slightly more general definition of the
coherent state. Introducing an arbitrary functional $f[R(a_1)]$ of the
remainder in Eq. (\ref{eqsi}), we define the coherent state to be
\begin{equation}
  \label{1}
  \mid z\rangle = \sum_{n=0}^{K} \left( z f[R(a_1)] \hat B_-^{-1}
  \right)^n  \mid 0\rangle 
\end{equation}
where the upper limit $K$ in the sum, depending on the nature (the
number of eigenstates) of the potential, can be either finite or
infinite. Note that one can define a new variable $z' = z f(a_1)$,
i.e. this generalization permits $z$ to be a function of the variables
$a_1,a_2,a_3, \cdots$. One can write Eq. (\ref{1}) explicitly as
\begin{eqnarray}
  \label{2}
   \mid z\rangle &=&  \mid 0\rangle + z \> \> 
   \frac{f[R(a_1)]}{\sqrt{R(a_1)}} \mid 1 \rangle + z^2
   \frac{f[R(a_1)]f[R(a_2)]}{\sqrt{R(a_2)[R(a_1)+R(a_2)]}} \mid 2
   \rangle \\
   &+& z^3
   \frac{f[R(a_1)]f[R(a_2)]f[R(a_3)]}{\sqrt{R(a_3)[R(a_2)+R(a_3)] 
[R(a_3)+R(a_2)+R(a_1)]}} 
   \mid 3 \rangle + \cdots
\end{eqnarray}
where $\mid n \rangle$ is the $n$-th excited state of the system: 
\begin{equation}
\mid n\rangle = \left[ \hat H^{-1/2}\hat B_+ \right]^n\mid 0\rangle\,.
\label{eqpsn}
\end{equation}
(cf. with Eq. (\ref{a2}). 

If the ground state is normalized, i.e. $\langle 0\mid 0\rangle = 1$,
then all the excited states given by Eq. (\ref{eqpsn}) are normalized
as well. If the number of energy eigenstates are infinite the coherent
state defined in Eq. (\ref{1}) is also an eigenstate of the operator
$\hat B_-$:  
\begin{equation}
\hat B_-\mid z\rangle = z f[R(a_0)] \mid z\rangle. 
\label{3}
\end{equation}
The additional condition
\begin{equation}
\left[ \hat B_- - z f[R(a_0)] \right] \frac{\partial}{\partial z}\mid
z\rangle =  f[R(a_0)] \mid z \rangle 
\label{eqbz}
\end{equation}
is also satisfied. 

\section{q-Coherent States for Shape-Invariant systems}

We now show that when $R(a_1)$ is given by Eq. (\ref{eqsc1}) 
and is subject to the scaling transformation given in Eq. (\ref{eqsc})
the coherent state of Eq. (\ref{1}) reduces to the standard q-coherent
states introduced in Refs. \cite{Biedenharn:1989jw} and
\cite{Macfarlane:dt}. In this case Eq. (\ref{1}) can be written in the
form 
\begin{equation}
  \label{4}
  \mid z\rangle = \sum_{n=0}^{\infty} \left( f[R(a_1)] f[R(a_2)]
    \cdots f[R(a_n)] \right) 
\frac{(1-q)^{n/2}}{\sqrt{(q;q)_n}} q^{-n(n-1)/4} 
\frac{z^n}{\sqrt{\left[R(a_1)\right]^n}} \mid n \rangle \,. 
\end{equation}
We now choose $ f[R(a_n)] =  R(a_n)$. The coherent state of
Eq. (\ref{4}) takes the form
\begin{equation}
  \label{5}
\mid z\rangle = \sum_{n=0}^{\infty} \frac{(1-q)^{n/2}}{\sqrt{(q;q)_n}}
q^{n(n-1)/4}  \sqrt{\left[R(a_1)\right]^n} z^n \mid n \rangle. 
\end{equation}
Using the normalized eigenstates given in Eq. (\ref{a2}), the above
equation can be written as
\begin{equation}
  \label{5a}
  \mid z\rangle = \sum_{n=0}^{\infty} \frac{(1-q)^n}{(q;q)_n} 
q^{n(n-1)/4} \left( z \sqrt{\left[R(a_1)\right]} \hat S_+ \right)^n
\mid 0 \rangle.  
\end{equation}
Using the q-exponential utilized in Ref. \cite{vinet} (see also
Refs. \cite{atakisi} and \cite{exton})
\begin{equation}
  \label{5b}
  E^{(\mu)}_q (x) = \sum_{n=0}^{\infty} \frac{q^{\mu n^2}}{(q;q)_n}
  x^n\,, 
\end{equation}
and the identity  $\sqrt{\left[R(a_1)\right]} \hat S_+ =  \hat B_+$,
the coherent state can be written as a generalized exponential:
\begin{equation}
  \label{5c}
   \mid z\rangle =  E^{(1/4)}_q \left( \frac{(1-q)}{q^{1/4}} z  \hat
     B_+ \right) \mid 0 \rangle \,.
\end{equation}
Finally introducing the variable
\begin{equation}
  \label{6}
  \zeta = \frac{\sqrt{(1-q)}}{\sqrt{q}} \sqrt{R(a_1)} z
\end{equation}
the coherent state can also be written as
\begin{equation}
  \label{7}
  \mid \zeta \rangle = \sum_{n=0}^{\infty}
  \frac{q^{n(n+1)/4}}{\sqrt{(q;q)_n}} \zeta^n  \mid n \rangle.
\end{equation}
The norm of this state is also given by a generalized exponential: 
\begin{equation}
  \label{8}
  \langle \zeta \mid \zeta \rangle = E^{(1/2)}_q \left( q^{1/4} \mid
    \zeta \mid^2 \right) . 
\end{equation}
Eq. (\ref{8}) is useful when writing path integrals in coherent-state
representation. 

\section{Completeness of q-coherent States and Ramanujan Integrals}

In this section we investigate the completeness relation for
q-analogue coherent states given in the previous section. One way to
obtain a decomposition of identity using q-analogue coherent states is
to use Jackson's formula for the q-differentiation and q-integration
\cite{jackson}. Such a completeness relation was introduced by the
authors of Ref. \cite{Gray:ea} (see also Ref.\cite{Bracken:dw}). In
the current work we introduce a completeness relation {\em without}
using q-integration. 

Our main result is that the decomposition of identity for the coherent
states of Eq. (\ref{7}) is given by 
\begin{equation}
  \label{9}
  I= \int \frac{d\zeta d\zeta^*}{2\pi i} \frac{1}{(-\log q)}
  \frac{1}{(-\mid \zeta \mid^2;q)_{\infty}}  \mid \zeta \rangle
  \langle \zeta \mid = \hat 1  
\end{equation}
which we prove in this section. (In the above relation $\hat 1$ is the
identity operator in the Hilbert space of the q-analogue harmonic
oscillator). 

We change of the variables $\zeta = \sqrt{t} e^{i \theta}$, using
Eq. (\ref{7}) write the coherent states in terms of the eigenstates of
the Hamiltonian, and perform the $\theta$-integration to obtain 
\begin{equation}
  \label{10}
  I= \frac{1}{(-\log q)} \sum_{n=0}^{\infty}
  \frac{q^{n(n+1)/2}}{(q;q)_n}  \mid n 
  \rangle \langle n \mid \times \int_0^{\infty} dt
  \frac{t^n}{(-t;q)_{\infty}} \,. 
\end{equation}
The last integral was evaluated by Ramanujan in an attempt to
generalize integral definition of the beta function \cite{raman}. (An
elementary proof is given by Askey \cite{askey1}). It is given by
\begin{equation}
  \label{11}
  \int_0^{\infty} dt \frac{t^n}{(-t;q)_{\infty}} =
  \frac{(q;q)_n}{q^{n(n+1)/2}} (-\log q) .
\end{equation}
Inserting Eq. (\ref{11}) into Eq. (\ref{10}) provides the proof of
Eq. (\ref{9}):
\begin{equation}
  \label{12}
  I =  \sum_{n=0}^{\infty} \mid n \rangle \langle n \mid = \hat 1. 
\end{equation}
As we mentioned above another decomposition of identity for q-coherent
states is given in Ref. \cite{Gray:ea}. In this reference a q-analogue
of the Euler's formula for $\Gamma (z)$ expressed as a q-integral is
utilized instead of the Ramanujan's integral,
Eq. (\ref{11}). Consequently the authors of Ref. \cite{Gray:ea}
obtained the resolution of identity for q-analogue harmonic
oscillators as a q-integral in contrast to our result, Eq. (\ref{12}),
where this resolution is expressed as an ordinary integral over
complex variables. 

\section{Conclusions}

In this paper we presented an overcompleteness relation for the
q-analogue coherent states as a complex integral. In arriving this
result we utilized shape-invariant algebraic techniques. Such a result
could be very useful in building coherent-state of path integrals
\cite{Fricke:1987dn} for q-analogue systems. Of course such path
integrals can also be defined over q-integrals. However, if one wishes
to approximate these path integrals using saddle-point approximations
or numerically evaluate them using Monte Carlo techniques ordinary
path integrals defined over complex variables present several
advantages over the q-integrals. Among these are the presence of a
continuous path to find stationary phases and the existence of 
positive-definite probabilities to do Monte Carlo integration. 

Coherent states for the ordinary harmonic oscillator has a dynamical
interpretation. If a forced harmonic oscillator in in its ground state
for $t=0$, it evolves into the harmonic oscillator coherent state. As
it was shown in Ref. \cite{Balantekin:1998wj} the coherent states
described here, in general, do not have such a simple dynamical
interpretation. However the evolution operator for the q-analogue of
the forced harmonic oscillator Hamiltonian
\begin{equation}
\hat h(t) = \hat B_+\hat B_- + f(t)\left[ {\rm
e}^{iR(a_1)t/\hbar}\hat B_+ +\hat B_- {\rm
e}^{-iR(a_1)t/\hbar}\right] \,,
\label{eqohf}
\end{equation}
where $f(t)$ is an arbitrary function of time, can be written as a path
integral where the integration path is given by the time-dependence of
the variable $\zeta$ of Eq. (\ref{9}) 
(cf. Ref. \cite{Fricke:1987dn} for the forced harmonic oscillator). 

Our decomposition of
identity can easily be generalized to multi-dimensional q-oscillators
and can be used to study for example the dynamics of $SL_q(n)$, which
can be constructed from such oscillators using a q-deformed
Levi-Civita tensor \cite{Arik:1993ey}. (For a review of various
applications of quantum groups see
e.g. Ref. \cite{Bonatsos:1999xj}). Coherent states for harmonic
oscillator representation of orthosymplectic superalgebras along with
their overcompleteness relations are given in
Refs. \cite{Balantekin:1987rp} and \cite{Balantekin:1988gw}. (For an
application of such representations see
Ref. \cite{Balantekin:1992qp}). Our techniques can be utilized to
determine the integration measures for the q-extensions of
superalgebras. 

Askey reports that Hardy mentions the Ramanujan integral of
Eq.(\ref{11}) to be new and interesting \cite{askey1}. Askey further
remarks that this integral is even more interesting than Hardy seems
to have thought it was and points out that it does not seem to have
arisen in applications very often. As we demonstrated this integral
seems to play a major role in building the Hilbert space for the
q-oscillator Hamiltonian.  

\section*{ACKNOWLEDGMENTS}

We are grateful to R. Askey for enlightening discussions and bringing
several references to our attention. 
This work was supported in part by the U.S. National Science
Foundation Grants No. INT-0070889 and PHY-0070161 at the University of
Wisconsin, and in part by the University of Wisconsin Research
Committee with funds granted by the Wisconsin Alumni Research
Foundation. M.A.C.R.\  acknowledges the support of Funda\c c\~ao de
Amparo \`a Pesquisa do Estado de S\~ao Paulo (Contract No.\
98/13722-2).  A.N.F.A. and M.A.C.R. acknowledge the support of
Conselho Nacional de Pesquisa (CNPq) (Contract No. 910040/99-0).
A.N.F.A. and M.A.C.R. thank to the Nuclear Theory Group at University
of Wisconsin for their very kind hospitality.

\end{document}